\documentclass[amsmath,nobibnotes,aps,superscriptadress,amssymb,pra,aps,showpacs,twocolumn]{revtex4-2}
\usepackage{amsmath,amsfonts,amssymb,amsthm,graphics,graphicx,epsfig,bbm}
\usepackage[colorlinks=true,citecolor=blue,linkcolor=blue,urlcolor=blue]{hyperref}
\usepackage[usenames]{color}
\usepackage{graphicx}
\usepackage{subfigure}
\usepackage{amsmath}
\usepackage{epsfig}
\usepackage{dcolumn}
\usepackage{bm}
\usepackage{color}
\usepackage{times}
\usepackage{epstopdf}
\usepackage{amssymb}
\usepackage{amstext}
\usepackage{latexsym}
\usepackage{hyperref}
\usepackage{amsfonts}
\usepackage{psfrag}
\usepackage{soul,xcolor}
\usepackage[normalem]{ulem}
\usepackage{dsfont}
\usepackage{physics}
\usepackage{footnote}
\usepackage{tikz}
\usepackage{blindtext}
\usepackage{bbm}
\usepackage{float}

\begin{document}

\title[Mitigating noise in digital and digital-analog quantum computation]{Mitigating noise in digital and digital-analog quantum computation}

\author{Paula Garc\'ia-Molina$^{1,2}$}
\email[Corresponding author: ]{\quad paula.garcia@iff.csic.es}
\author{Ana Martin$^{1,3}$}
\author{Mikel Garcia de Andoin$^{4,1,3}$}
\author{Mikel Sanz$^{1,3,5,6}$}

\affiliation{$^1$Department of Physical Chemistry, University of the Basque Country UPV/EHU, Apartado 644, 48080 Bilbao, Spain}
\affiliation{$^2$Institute of Fundamental Physics, IFF-CSIC, Calle Serrano 113b, 28006 Madrid, Spain}
\affiliation{$^3$EHU Quantum Center, University of the Basque Country UPV/EHU, 48940 Leioa, Spain}
\affiliation{$^4$TECNALIA, Basque Research and Technology Alliance (BRTA), 48160 Derio, Spain}
\affiliation{$^5$IKERBASQUE, Basque Foundation for Science, Plaza Euskadi 5, 48009 Bilbao, Spain}
\affiliation{$^6$Basque Center for Applied Mathematics (BCAM), Alameda Mazarredo 14, 48009 Bilbao, Spain}


\begin{abstract}
Noisy Intermediate-Scale Quantum (NISQ) devices lack error correction, limiting scalability for quantum algorithms. In this context, digital-analog quantum computing (DAQC) offers a more resilient alternative quantum computing paradigm that outperforms digital quantum computation by combining the flexibility of single-qubit gates with the robustness of analog simulations. This work explores the impact of noise on both digital and DAQC paradigms and demonstrates DAQC’s effectiveness in error mitigation. We compare the quantum Fourier transform and quantum phase estimation algorithms under a wide range of single and two-qubit noise sources in superconducting processors. DAQC consistently surpasses digital approaches in fidelity, particularly as processor size increases. Moreover, zero-noise extrapolation further enhances DAQC by mitigating decoherence and intrinsic errors, achieving fidelities above 0.95 for 8 qubits, and reducing computation errors to the order of $10^{-3}$. These results establish DAQC as a viable alternative for quantum computing in the NISQ era.

\end{abstract}

\maketitle
\section*{Introduction}

The origin of quantum computing dates back to 1980 when Benioff \cite{Benioff1980} suggested using quantum systems for traditional (reversible) computation, implementing a Turing-machine equivalence to unitary evolution. Simultaneously, Manin \cite{Manin1980}, and a couple of years later, Feynman \cite{Feyn1982}, pointed out the convenience of quantum simulators to reproduce quantum problems. It was not until 1985 when Deutsch \cite{Deutsch1985} proposed the idea of a quantum gate, which led to the gate-based quantum computation model known as the digital quantum computing (DQC)\footnote{Note that DQC also stands for other common terms in quantum computing, such as differentiable quantum circuits or distributed quantum computers, among others. In this text, DQC only stands for digital quantum computing.} paradigm.

DQC uses digital pulses to perform quantum operations on the qubits, represented by single-qubit gates (SQGs), two-qubit gates (TQGs), and measurements. While DQC is the most widespread paradigm, other alternatives are based on analog resources. One of the most promising ones is called adiabatic quantum computation or quantum annealing \cite{Kadowaki1998, Farhi2000, Albash2018}. Besides these paradigms, digital-analog quantum computing (DAQC) merges the digital and analog paradigms, combining the adaptability of DQC with the robustness of analog resources \cite{Parra2018}. This universal paradigm consists in applying digital and analog blocks to implement quantum circuits in a quantum computer. The digital steps consist of SQGs, while analog blocks are constructed via the time evolution of the natural interacting Hamiltonian provided by the quantum processor. Several works employ this paradigm to simulate quantum many-body systems \cite{LPSS2018, Celeri2021} and to implement long-depth quantum algorithms \cite{MLSS2020, Headley2020}.

Since the inception of quantum computing, much effort has been put into developing the area. There is a great variety of hardware backends suitable for the performance of quantum computation, such as trapped ions \cite{Cirac1995}, neutral atoms~\cite{Wintersperger2023} or superconducting circuits. The latest is one of the most promising quantum computer construction techniques. Solid-state electric circuits use Josephson tunnel junctions, a highly nonlinear and non-dissipative circuit element at low temperatures \cite{superconducting}. 

Noise is undoubtedly the main drawback in the scalability of quantum computing. Indeed, quantum computers cannot be completely isolated from the environment, and the interactions performing quantum operations cannot be fully controlled, leading to decoherence, control errors, and crosstalk. Due to the limitations imposed by these error sources, considerable effort has been made to analyze them and propose quantum protocols more resistant to noise or capable of mitigating their effects\ \cite{Zhang2019, Sarovar2020}. The most promising long-term solution for the scalability of quantum computers is the use of quantum error correction techniques\ \cite{Shor1995,Bennett1996}. However, the current Noisy Intermediate-Scale Quantum (NISQ)\ \cite{Preskill2018,Bharti2021} era technology is not sufficient to implement quantum error correction to eliminate the undesired effect of noise sources in them.

A possibility to reduce the effect of noise sources in the era of NISQ devices is to adapt the quantum computing paradigm. Although the gate-based approach is the most extended paradigm, it is highly affected by noise. However, the effect is different depending on the type of gate. Indeed, it is almost negligible in SQGs since they are fast and highly controllable. The length of TQGs exposes them to decoherence, which makes these gates very sensitive to noise, together with the errors introduced by the indirect control and the crosstalk among qubits. DAQC emerges as a sensible alternative quantum paradigm to overcome this problem within the limitations of the NISQ era. This paradigm employs the natural interaction of the quantum processor to perform entangling operations and the running time of this Hamiltonian as an extra resource \cite{Gonzalez2021}.
Consequently, the crosstalk among qubits and the control errors associated with switching on and off interactions are no longer a source of noise but a resource, turning the problem of noise in TQGs into an advantage. Indeed, this approach---and more concretely, its banged version---has been numerically demonstrated to be more advantageous under the effect of specific experimental noise sources when compared to DQC \cite{MLSS2020}. 

Ref.~\cite{MLSS2020} considered a limited type of noise source, which may not represent all the sources present in current devices. A more comprehensive comparison of the performance of DAQC and DQC under common noise sources in superconducting quantum computers is necessary. In addition, Ref. \cite{MLSS2020} shows that, while DAQC may reduce noise impact compared with DQC, the decrease in fidelity with the number of qubits makes it challenging to derive significant results, similar to most DQC applications.  Quantum error mitigation \cite{Endo2018,Kandala2019, Endo2021} is a promising tool to reduce noise-related errors in NISQ processors. Expanding error mitigation to the DAQC paradigm could enhance its performance in the NISQ era.

This article presents an extension of error mitigation to the DAQC paradigm, providing a detailed comparison of the performance of DQC and DAQC under a wide range of noise sources---including thermal decoherence, bit-flip, measurement errors, and control errors. The quantum Fourier transform (QFT)---an essential quantum subroutine with applications in multiple quantum algorithms \cite{Shor1996,MCRMCLOSS2021,GRSM2021,HHL2009,LMR2014}---acts as a first benchmark. The numerical study covers quantum processors of up to 6 qubits, comparing the fidelity of the resulting state of the different paradigms to the exact solution. The results conclude that the banged DAQC paradigm better tolerates the most relevant noise sources in superconducting quantum, especially as the number of qubits scales up. This may allow a more accurate implementation of quantum algorithms in the NISQ era than DQC. These conclusions also extend to more complex algorithms, such as phase estimation (QPE). Given a unitary operator $U$ with an eigenvector $|u\rangle$ with eigenvalue $e^{2\pi \varphi}$, this algorithm estimates the unknown phase $\varphi$. It hence has multiple applications, such as in the Shor's\ \cite{Shor1995} and Harrow-Hassidim-Lloyd algorithms\ \cite{HHL2009}. This algorithm includes the inverse QFT besides some conditional rotations and SQGs, so we expect it to pose a more significant challenge to the paradigms due to the greater effect of decoherence and other noise sources associated with the greater number of quantum gates. However, despite these larger noise sources, we show how the DAQC paradigm can approximate the exact solution with fidelities overcoming those of DQC. To validate these statements, we study in detail a QPE circuit for 5 qubits and perform extensive simulations in which we investigate the effect of each error source independently. Despite the better noise resilience of DAQC compared with DQC, the results show that the error increase with qubit scaling avoids extracting useful information from the solution of the algorithms. To avoid this, we extend zero noise extrapolation\ \cite{Li2017,Temme2017} to the DAQC paradigm. This technique successfully removes the noise of decoherence---one of the main noise sources in quantum devices---, together with the one inherent to the banged DAQC paradigm, achieving fidelity above 0.95 for 8 qubits and errors in the computation of expectation values of order $10^{-3}$.

\section*{Results}

\section{Implementation of a noise model using the DQC and DAQC paradigms}

In this section, we illustrate the noise model described in section \hyperref[sec: model]{Theoretical description of the effect of noise sources in quantum systems} with the QFT and QPE algorithms. 

\subsection{Quantum Fourier transform} \label{sec: QFT}

In Ref. \cite{MLSS2020}, it is shown that the N-qubit QFT can be written as a series of unitary gates of the form
\begin{equation} \label{eq: U_QFT}
U_{\text{QFT}}=\left[\prod_{m=1}^{N-1}U_{\text{SQG},m}U_{\text{TQG},m}\right]\cdot U_{H,N},
\end{equation}
where
\begin{align}
U_{\text{SQG},m}&=\exp\left[i\sum_{k=2}^{N-(m-1)} \theta_k\left(\mathbbm{I}-Z^{(k+m-1)}-Z^{(m)}\right)\right]\times\nonumber\\
&\times \exp\left[ \frac{i\pi}{2}\left(\mathbbm{I}-\frac{Z^{(m)}+X^{(m)}}{\sqrt{2}}\right)\right], \label{eq:U_SQG}\\
U_{\text{TQG}}&=\exp\left(i\sum_{c<k}^N\alpha_{c,k,m}Z^{(c)}\otimes Z^{(k)}\right),\label{eq:H_ZZ}\\
U_{H,m}&=\exp\left(\frac{i \pi}{2}\left[\mathbbm{I}-\frac{(Z^{(m)}+X^{(m)})}{\sqrt{2}}\right]\right),\label{eq:U_Hm}\\
\theta_k&=\frac{\pi}{2^{k+1}}\qquad\text{and}\qquad\alpha_{c,k,m}=\delta_{c,m}\frac{\pi}{2^{k-m+2}}\label{eq:theta_alpha},
\end{align}
where the superindices in brackets determine which qubit the gate acts on. We compute the tensor product with the identity to obtain the total unitary until dimension $2^N$ is reached.

The entangling gates of the algorithm are described by Eq.~\eqref{eq:H_ZZ}, which is the unitary representation of an inhomogeneous all-to-all (ATA) two-body Ising Hamiltonian. The DQC paradigm decomposes the entangling gate into fixed $\pi/4$ phase inhomogeneous ATA two-body Ising Hamiltonians as
\begin{align} \label{eq:ZZ gates}
e^{i\alpha_{c,k,m}Z ^c Z ^k}&=e^{i\frac{\pi}{4}Y^c}e^{i\frac{\pi}{4}Z^c Z^k}e^{i\alpha_{c,k,m} Y^c} X^k  \nonumber\\ &\times e^{i\frac{\pi}{4} Z^c Z^k}X^k e^{-i\frac{\pi}{4} Y^c},
\end{align}
where the phase $\alpha_{c,k,m}$ is the coefficient in Eq. \eqref{eq:theta_alpha}.

To obtain a fair comparison of the performance of DQC, sDAQC, and bDAQC for the implementation of the QFT, we choose a series of initial states from the family of states $\ket{\psi_0}=\sin\beta\ket{\text{W}_N}+\cos\beta\ket{\text{GHZ}_N}$, where $\beta \in [0, \pi]$. The comparison is performed for the 3-, 5- and $6-$qubit QFT so we can observe the effect of the noise sources as the number of qubits increases.

The noise that we consider encloses the most relevant noise sources in superconducting quantum computers: bit-flip, decoherence, measurement error, and control-related errors such as magnetic field fluctuations. However, the difference in the implementation of the entangling blocks in DQC and DAQC leads to a different experimental error associated with them, which is modeled as a Gaussian phase noise on the application time of the TQGs in DQC and the analog blocks in DAQC. 

The figure of merit used for the comparison is the fidelity
\begin{equation}\label{eq:fidelity}
    F(\rho_{\text{ideal}},\rho_{\text{noisy}})=\left[\text{tr}\left(\sqrt{\sqrt{\rho_{\text{ideal}}}\rho_{\text{noisy}}\sqrt{\rho_{\text{ideal}}}}\right)\right]^2,
\end{equation}
where $\rho_{\text{ideal}}$ is the ideal state after the application of the QFT, while $\rho_{\text{noisy}}$ is the resulting state when the noise model is implemented. The fidelity verifies $0\leq F(\rho_{\text{ideal}},\rho_{\text{noisy}})\leq 1$, and the larger the values the more similar are $\rho_\text{ideal}$ and $\rho_\text{noisy}$.

\begin{figure}
\centering
\includegraphics[width=0.48\textwidth]{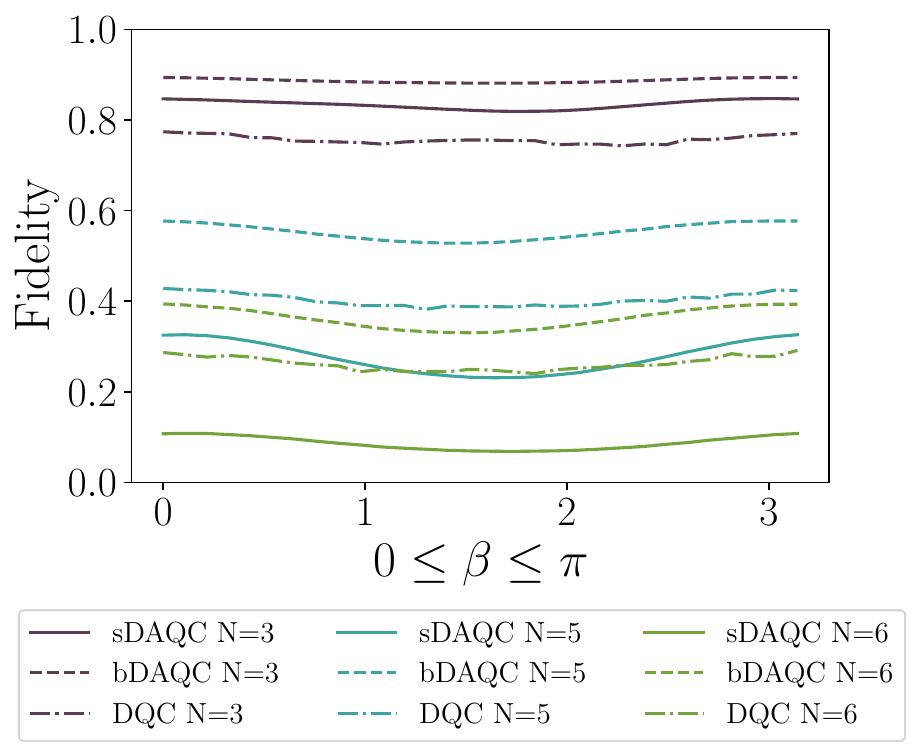} 
\caption{{\bf Fidelity of the implementation of the 3-, 5- and 6-QFT algorithm using the DQC, sDAQC, and bDAQC paradigms.} The algorithm is applied to the family of states $\ket{\psi_0}=\sin\beta\ket{W_N}+\cos\beta\ket{GHZ_N}$ .DQC stands for digital quantum computing, while sDAQC and bDAQC stand for stepwise and banged digital-analog quantum computing. The colors purple, teal, and green represent the number of qubits of the quantum circuits, which are $N=3,5,6$, respectively. The linestyles determine the paradigms: solid for sDAQC, dashed for bDAQC, and dash-dot for DQC. The highest fidelity is obtained for the bDAQC protocol. DAQC presents a better tolerance to decoherence than DQC due to the shorter time of application of the analog blocks with respect to TQGs. Moreover, the effect of bit-flip error in bDAQC is smaller than on sDAQC as the analog interaction is always on, so the bit-flip only affects single-qubit rotations (SQRs) in bDAQC, as opposed to sDAQC in which all the qubits of the system are subjected to it when an analog block is applied.}
\label{Fig:simulation}
\end{figure}

The values of the parameters used to simulate the different noise sources are chosen to model realistic NISQ superconducting devices. For the experimental errors described in Ref.~\cite{MLSS2020}, we select the values proposed by the authors. The magnetic field noise acting on single-qubit gates is given by a uniform probability distribution $\mathcal{U}(1-\text{SQGN},1+\text{SQGN})$, with $\text{SQGN} = 0.0005$. For the entangling blocks, for both DQC and DAQC, the noise is modeled by a random variable from a Gaussian distribution $\mathcal{N}(0,\sigma)$. The standard deviation value $\sigma$ depends on the paradigm used. For DQC, the standard deviation is TQGN$=0.2000$, while for DAQC, we have sABN$=0.0200/g_0$ and bABN$=0.0100/g_0$, for sDAQC and bDAQC, respectively. The parameter $g_0$ is the coupling constant of the homogeneous ATA two-body Ising Hamiltonian of the analog blocks, and it is equal to $10 \text{ MHz}$. On the other hand, for the incoherent noise models introduced in Section \hyperref[sec: model]{Theoretical description of the effect of noise sources in quantum systems}, we have a bit-flip error with probability $p_\text{b-f}=0.005$ and a measurement error with probability $p_\text{meas}=0.01$. Notice that measurement error is currently a great technological limiting factor for quantum algorithms run in superconducting quantum processors. Finally, for the decoherence channel, the relaxation time $T_1$ is $50\ \mu \text{s}$, with the ground state thermal population $p=0.35$. The length of the SQGs is $\Delta t_\text{SQG}= 1/(100g_0)$, while for the two-qubit interaction, it is $\Delta t_\text{TQG}= (1+\varepsilon)100\Delta t_\text{SQG}\pi/4$ for TQGs, and for the analog blocks, it is the one obtained from the decomposition of the interaction. 

The simulation using the previous values of the parameters is depicted in Fig. \ref{Fig:simulation}, averaged over 1000 repetitions to obtain a sufficiently good statistical sample. We can conclude that the bDAQC protocol performs better when the noise sources considered by our noise model are considered. Although the lowest fidelity results are obtained for the ideal case for bDAQC \cite{MLSS2020}, in the noisy simulation, we observe that it shows the best error tolerance. The reason is that the error due to turning the interaction on and off overcomes the inherent error of the bDAQC introduced by always having the interaction on. Moreover, the fact that the interaction is never switched off also makes bDAQC only affected by the bit-flip error for the SQGs, so no bit-flip error is introduced due to the entangling blocks as it does in both DQC and sDAQC. For the chosen set of parameters, the total time of application of the QFT algorithm using the DAQC paradigm is shorter than the total time using DQC. Therefore, the DAQC paradigm is less affected by decoherence than the DQC paradigm. The measurement error is common to the three paradigms, so its effect on the fidelity is reflected as a shift in all the values. The performance of the QFT algorithm decreases fast with the number of qubits due to the increase in depth of the algorithm, as the number of gates is of order $\mathcal{O}(N^2)$, where $N$ stands for the number of qubits. Thus, under a fair comparison considering the experimental values and the same noise sources, bDAQC obtains the highest value of fidelity.

Although we have chosen the QFT algorithm for our comparison, it allows us to extend the results to any general quantum algorithm, as the noise sources considered are independent of the algorithm implemented. In addition, if the scaling of the number of quantum gates with the number of qubits is slower than in the QFT, better fidelity results are expected.

\subsection{Quantum phase estimation}

To extend our study of the effect of noise sources in DQC and DAQC paradigms, we propose the QPE algorithm \cite{Nielsen2000}. This algorithm estimates the phase $\varphi$ that corresponds to the eigenvalue $e^{2\pi i \varphi}$ of the eigenvector $|u\rangle$ of a unitary operator $U$, where $\varphi$ is unknown. In this algorithm, we use two quantum registers. In the first one, we use $t$ qubits that determine the accuracy in the approximation of $\varphi$. The second register is used to encode the eigenvector $|u\rangle$. Then, we initialize the first register by applying a Hadamard gate on each qubit. Then, we apply a series of controlled rotations on the second register with $U$ exponentiated to increase the powers of two. After that, we use the inverse Fourier transform on the first register, which allows us to recover an approximation of $\varphi$, hence the eigenvalue we were looking for.

In the previous section (Subsec.\ \hyperref[sec: QFT]{Quantum Fourier transform}), we studied how the QFT's precision scales with the number of qubits using DQC, sDAQC, and bDAQC under noisy circumstances. We observed that for a growing number of qubits, the DAQC paradigm outperforms DQC. Therefore, as the inverse QFT is part of the QPE algorithm, we would expect it to have a similar performance. We propose pushing these paradigms' limits by obtaining a concrete eigenvalue via the QPE algorithm. This algorithm poses a more significant challenge for the performance of DQC, as it includes a larger number of controlled two-qubit operations, which are the primary noise sources in current superconducting devices. Thus, we expect that using the analog evolution of DAQC will allow us to obtain higher fidelities than the digital approach. 

\begin{figure}[t]
\centering
\includegraphics[width = 0.48\textwidth]{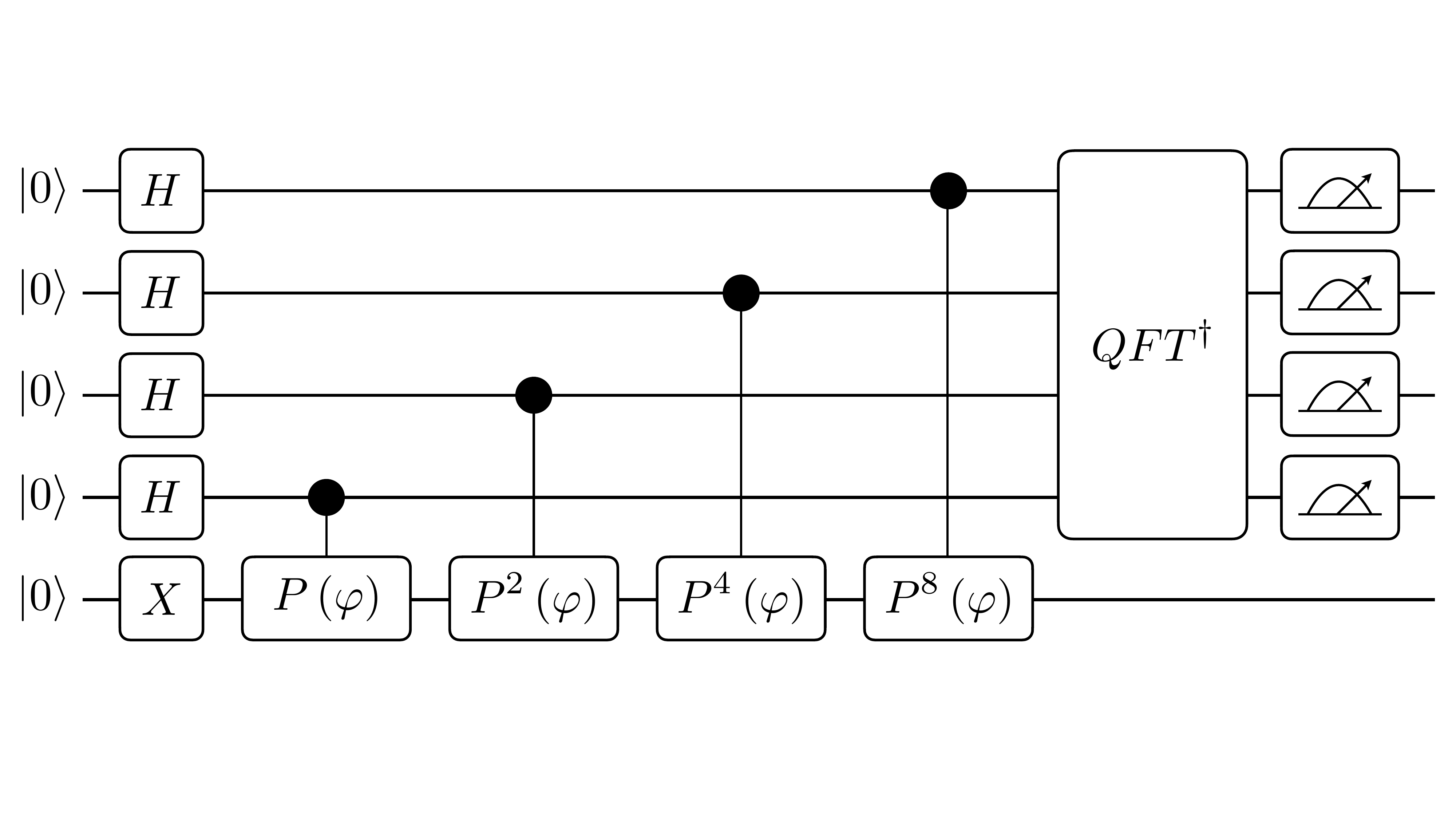}
\caption{\textbf{QPE algorithm.} QPE algorithm for the single qubit unitary $P\left(\varphi\right)$ from Eq.~\eqref{eq: QPE_matrix}, such that $P(\varphi)|1\rangle=e^{i2\pi\varphi}|1\rangle$, with 4 anciliary qubits. Here, the selected value of $\varphi$ is $1/3$, and so, it is the expected output of the circuit.}
  \label{Fig:qpe}
\end{figure}

Given the unitary operator $P\left(\varphi\right)$,
\begin{equation} \label{eq: QPE_matrix}
    P\left(\varphi\right) = \begin{pmatrix} 1 & 0 \\
                        0 & e^{i2\pi \varphi}
        \end{pmatrix},
\end{equation}  
we aim at obtaining $\varphi$ such that $P(\varphi)|u\rangle=e^{i2\pi\varphi}|u\rangle$. We choose a 4-qubit register to estimate the phase $\varphi$ that corresponds to the eigenvector $|u\rangle = |1\rangle =(0 \quad 1)^T$. The complete QPE circuit is depicted in Fig.\ \ref{Fig:qpe}.

We follow a different approach to implement this circuit from the one for the QFT. In the previous algorithm, we tailored each quantum gate to obtain the best possible performance of the methods, for example, by combining and applying all quantum gates simultaneously on the same pulse, as described in Ref.\ \cite{MLSS2020}. However, in practical applications, quantum operations are reduced to a basis of quantum gates, each with a corresponding pulse, and all operations need to be rewritten in terms of them. Then, to apply this algorithm, we propose to choose a concrete set of quantum gates based on the ones in a real superconducting quantum computer. We choose the universal set $\lbrace R_x(\theta), R_z(\theta), \text{CNOT}\rbrace$, and construct a tool to map any DQC circuit in the basis gates to the corresponding decomposition in the DAQC paradigm.

The developed tool simulates the noise sources in the quantum computer, similar to the application to the QFT in Subsec.\ \hyperref[sec: QFT]{Quantum Fourier transform}, but uses a more general scheme capable of implementing various algorithms. Moreover, it also considers the effect of residual signals in neighboring qubits when applying the pulses for a SQG (crosstalk), effectively applying the bit-flip error to all qubits after each digital block. See the Code Availability section for access to the general tools developed due to this work.

We have used noise parameters based on state-of-the-art quantum computers in our simulations. In this case, we resort to approximated values of the execution times for the basis gates of IBM quantum computers, with $t_{R_x}=10\text{ns}$, $t_{R_z}=1\text{ns}$, and $t_{\text{CNOT}}=300\text{ns}$, which show that the execution time for TQGs is one or two orders of magnitude larger than the one of SQGs, which is consistent to each great effect to noise due to decoherence. The approximated error for these gates is of $\text{TQGN}=0.08$, with $\text{CNOT}=e^{-isH_\text{CNOT}}$, where $s \in \mathcal{N}(\pi/2,\text{TQGN})$. This error has also been estimated from real IBM devices. We choose a 0.0001 bit-flip probability for SQGs and the same value as before for the error measurement, $p_\text{meas}=0.01$. Regarding decoherence, we consider a $T_1$ time of $50\mu$s with the ground state thermal population $p=0.35$. Finally, for the experimental errors defined in Ref.\ \cite{MLSS2020} we use $\text{SQGN}=0.0005$, $\text{sABN}=0.002$, and $\text{bABN}=0.001$. The coupling strength in these simulations has been chosen to be $g_0=1$ (units of $g_0$).

\begin{figure}
\centering
\includegraphics[width=0.48\textwidth]{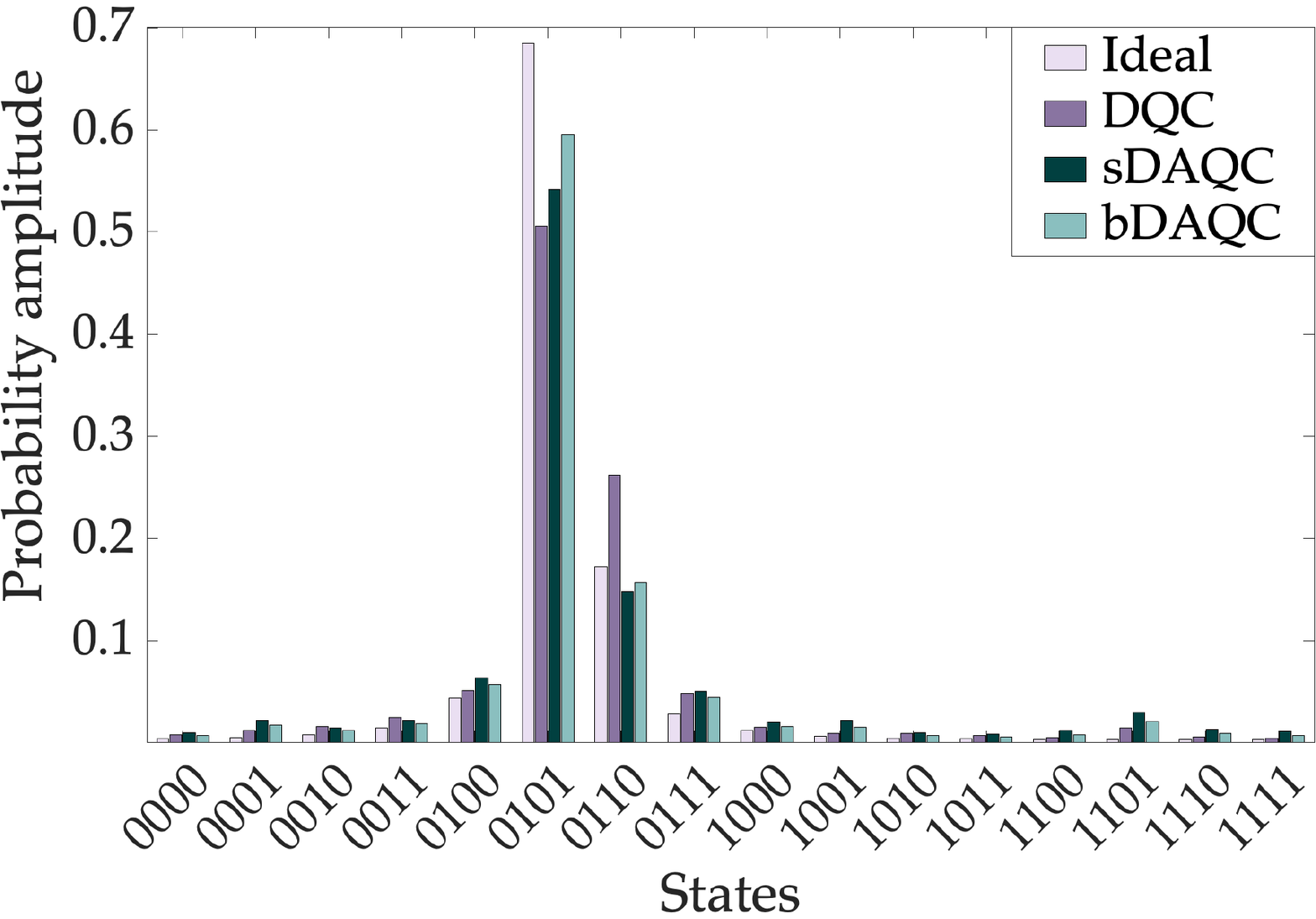} 
\caption{{\bf QPE results for DQC, sDAQC, and bDAQC.} Probability amplitudes of the resulting states of the QPE algorithm for the considered paradigms and noise sources for $\varphi = 1/3$ and the unitary operator in Eq.\ \eqref{eq: QPE_matrix}.}
\label{Fig: QPE}
\end{figure}

We use these parameters to estimate the phase $\varphi$ such that $P(\varphi)|u\rangle=e^{i2\pi\varphi}|u\rangle$, choosing $|u\rangle$ to be equal to $|1\rangle =(0 \quad 1)^T$. In Fig. \ref{Fig: QPE} we observe that under ideal circumstances, the higher probabilities are those of the $0101$ and $0110$ states, which lead to the approximations of $\varphi$, $\varphi_{0101} = 5/16 = 0.3125$, $\varphi_{0110} = 6/16 = 0.375$, with a relative error of $6.25 \% $ and $12.5\%$ respectively, with respect to the exact result $\varphi = 1/3$. We observe that we can recover the maximum probability states with DQC and DAQC even under noisy circumstances. If we calculate the weighted mean of the estimated phase, we obtain an estimate of phase of $\varphi_\text{bDAQC}=0.348\pm0.119$ for bDAQC, $\varphi_\text{sDAQC}=0.360\pm0.107$ for sDAQC and $\varphi_\text{DQC}=0.347\pm0.111$ for DQC. Using this figure of merit, we conclude that the results for bDAQC and DQC are practically identical. However, suppose we use majority voting as the estimation strategy. In that case, we can see that bDAQC outperforms the rest of the circuits as we reach the closest approximation ($\varphi=5/16$) faster than with the other approaches.

\begin{figure*}
    \centering
    \includegraphics[width=\linewidth]{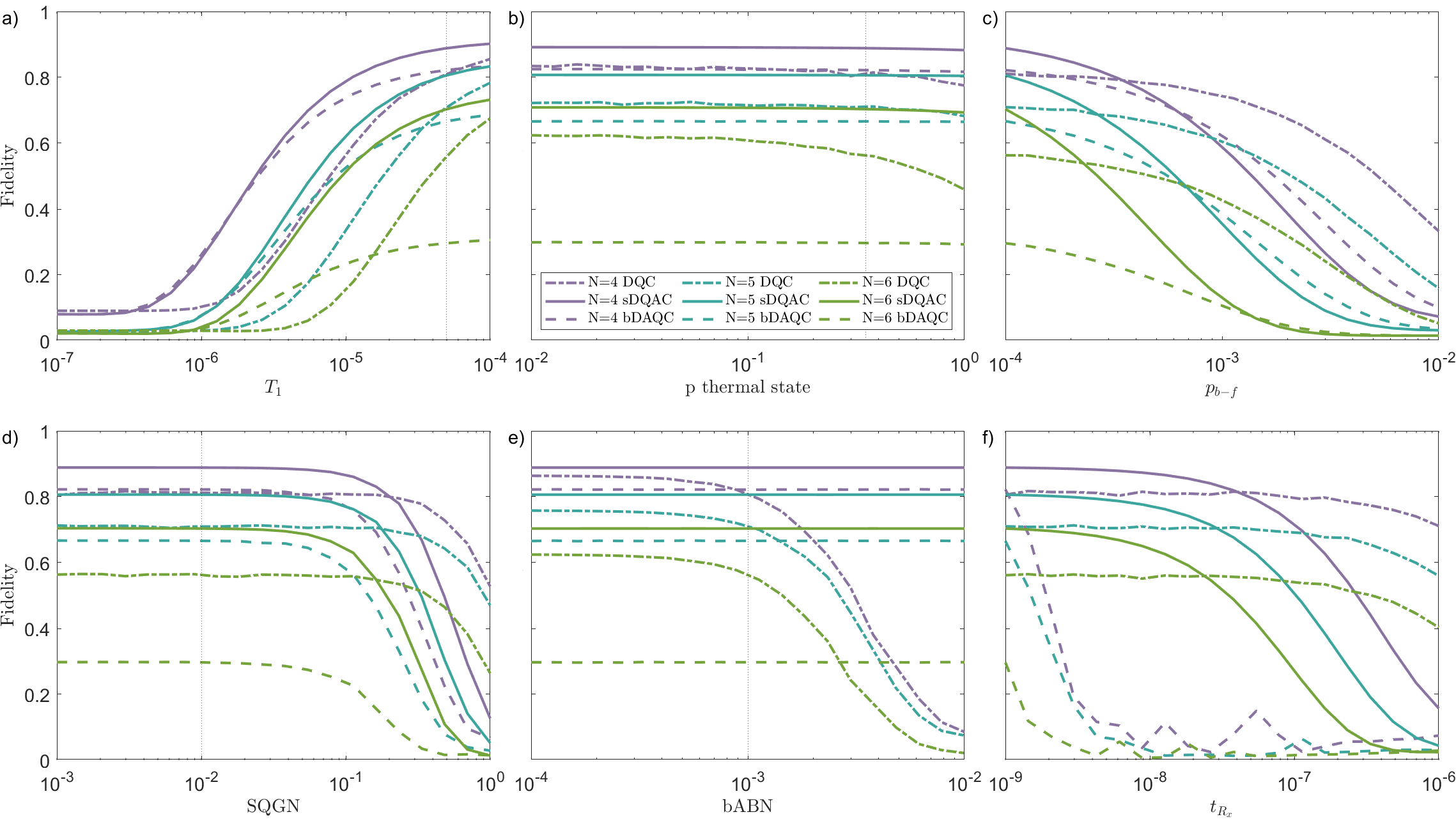}
    \caption{\textbf{Fidelity of the noisy implementation for the QPE.} The figure shows the results for the DQC (dot-dashed line), sDAQC (solid), and bDAQC (dashed) for circuits with 4 (violet), 5 (blue), and 6 (green) qubits. The unitary employed is the same as in Eq.~\eqref{eq: QPE_matrix}. The sDAQC and bDAQC circuits have been compiled using the same step-by-step construction as previously employed for the QPE. We set the noise parameters used in the main text for the simulations, and we only vary one of them. In (a) we vary the $T_1$ time, in (b) the thermal state probability, (c) $p_{b-f}$, (d) the SQG noise, (e) bABN and (f) the $R_x$ gate time. We show the average fidelity over 100 different simulations for each data point. The vertical lines show the value of the noise parameter employed in the main text. The fidelity is calculated using Eq.~\eqref{eq:fidelity} and by considering all qubits in the system.}
    \label{fig:benchmark}
\end{figure*}

Additionally to this detailed analysis of the performance for a single instance of QPE, we perform a more general benchmark for different system sizes and error parameters. The results are shown in Fig.~\ref{fig:benchmark}. First, let us study the decoherence error, determined by the thermal relaxation time $T_1\in [10^{-7}\text{s},10^{-4}\text{s}]$ and the ground state thermal population $p\in[10^{-2},1]$, respectively presented in Figures\ \ref{fig:benchmark}(a)-(b). In both cases, DAQC is more resilient to these errors than DQC, as the total circuit time for DQC is higher. Regarding the bit-flip $p_{b-f}\in[10^{-4},10^{-2}]$ Fig.~\ref{fig:benchmark}(c) and single-qubit gate error $\text{SQGN}\in[10^{-3},1]$ Fig.~\ref{fig:benchmark}(d) DAQC performs worse than DQC due to its higher single qubit gate count. sDAQC loses fidelity faster than bDAQC due to the assumption that the bDAQC protocol neglects the control errors of turning on and off the Hamiltonian, which could induce bit-flips. The parameters $\text{sABN},\text{bABN}$ and $\text{TQGN}$ determine the errors related to the interaction Hamiltonian and the two-qubit gate errors. Similarly to the previous simulation, we fix the relation $\text{sABN}=2\text{bABN}$ and $\text{TQGN}=80\text{bABN}$. As shown in Fig.~\ref{fig:benchmark}(e), for DAQC the errors in the analog blocks average out at the end of the circuit, avoiding affecting the fidelity. However, DQC does suffer from larger errors in the two-qubit gates. In Fig.~\ref{fig:benchmark}(f), the gate application time ranges from $1\text{ns}$ to $1\mu s$ to apply an $X$ rotation gate and one order of magnitude less time to apply a $Z$ rotation, $t_{R_z}=t_{R_x}/10$. DAQC is highly sensitive to slower SQGs; in sDAQC, as the total circuit time increases, the fidelity is more affected by decoherence. On top of this error, bDAQC longer gates increase the banging error to the point at which we very rapidly lose all information from the state. In general, bDAQC is more robust to the other kinds of errors considered and mitigating the banging error would reduce its main drawback. This highlights the need for a banging error mitigation technique. We address this issue in section\ \hyperref[sec: error mitigation]{Error mitigation in DAQC}.

\section{Error mitigation in DAQC} \label{sec: error mitigation}
The effect of noise sources limits the performance of current NISQ processors \cite{Preskill2018}. Quantum error mitigation techniques emerge to deal with these effects within the NISQ era successfully \cite{Endo2018,Kandala2019,Endo2021}. Some of the latest proposals of error mitigation techniques apply the quasi-probability method stochastically to analog, DAQC, or DQC platforms and efficiently can eliminate errors \cite{Endo2019,Sun2021}, or suppress algorithmic errors such as Trotter error \cite{Hakoshima2021}. Therefore, to show the actual performance of DAQC, it is necessary to show how to adapt quantum error mitigation techniques to this paradigm. We will demonstrate that these techniques can be successfully modified to cancel the intrinsic error associated with bDAQC. 

We apply the zero noise extrapolation technique\ \cite{Li2017,Temme2017}, based on the extrapolation of the noisy results to the zero noise limit. We consider only the effect of decoherence in the system and run the bDAQC QFT circuit for the initial state $\ket{\psi_0}=\sin\pi/4\ket{\text{W}_8}+\cos\pi/4\ket{\text{GHZ}_8}$, for different values of the coupling constant $g_j, \ j=0,\dots,n_g$. This way, we modify the total time of the application of the circuit as $t_{j_{\alpha}} \propto g_j^{-1}$ and hence, the effect of decoherence. However, the variation of $g_j$ also affects the intrinsic error of the bDAQC paradigm. To avoid this effect, we choose $\Delta t_{\text{SQG}_{i,j}} = b_i/g_j$, as for a given $g_j$ the time of the digital block is defined to be inversely proportional to the coupling constant. Then, for a certain number of values of $g_j$, it is possible to extrapolate the fidelity results to the zero decoherence limit-zero total time, removing the error due to decoherence.

\begin{figure}
\centering
\includegraphics[width=0.4\textwidth]{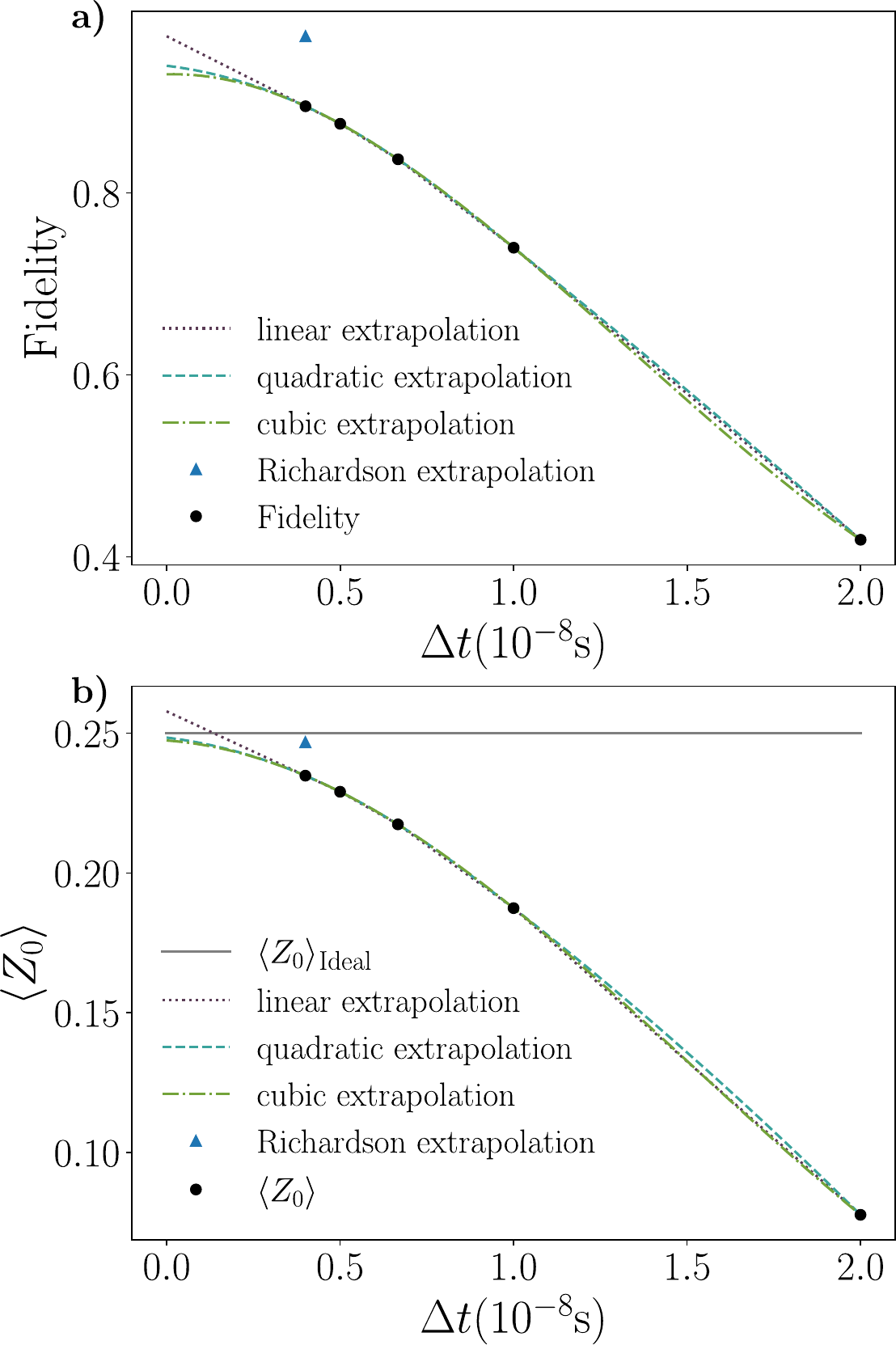} 
\caption{{\bf Results of the error mitigation.} (a) Fidelity, and (b) ${\langle Z_0 \rangle}$. The black dots represent the fidelity simulation values and ${\langle Z_0 \rangle}$.  The colors and linestyles represent the different extrapolations: purple dotted line for linear extrapolation, teal dashed line for quadratic extrapolation, green dash-dot line for cubic extrapolation, and blue triangle for Richardson extrapolation. The grey solid line in Figure (b) represents the ideal expectation value ${\langle Z_0 \rangle_\text{Ideal}}$. The simulations were made for the $8$-qubit QFT circuit with $T_1 = 50 \ \mu$s, $p=0.35$ and $g_0 = 1$ MHz.}
\label{Fig: error mitigation}
\end{figure}

Moreover, it is also possible to extend this technique to eliminate the intrinsic error introduced by bDAQC. To do that, we repeat the previous procedure for $i = 0, ..., n_t$ values of $\Delta t_{\text{SQG}_{i,j}} = b_i/g_j$, employing the zero decoherence values to perform the extrapolation to the zero $\Delta t_\text{SQG}$ limit. We choose an 8-qubit QFT circuit for applying this technique, as the number of qubits and quantum gates is challenging enough for state-of-the-art quantum devices. We obtain the fidelity of the ideal circuit, both without the error due to decoherence and bDAQC (Fig.\ \ref{Fig: error mitigation}a). The linear extrapolation gives the best fidelity, approximately 0.9723, which means that the error-mitigated state is an excellent approximation of the ideal one. The same result is obtained for Richardson extrapolation; in this case, the best result is obtained for the first-order approximation.

However, obtaining the quantum state after applying a quantum circuit is not always possible, as it is a costly operation due to the use of tomography. Thus, to verify the performance of error mitigation for bDAQC, it is necessary to choose a second figure of merit that can easily be computed from the measurements of a quantum circuit. We choose the expectation value of the observable $Z_0 = Z\otimes \mathbb{I}\otimes \dots \otimes \mathbb{I}$. The results are shown in Fig.\ \ref{Fig: error mitigation}b. The best approximation of $\langle Z_0 \rangle$ is approximately 0.2500, computed using Richardson extrapolation. The error $\varepsilon = | \langle Z_0 \rangle-\langle Z_0 \rangle_\text{approx} | = 0.0016$ shows that we can obtain a highly accurate approximation of the ideal expectation value from the noisy one.

\begin{figure}
\centering
\includegraphics[width=0.4\textwidth]{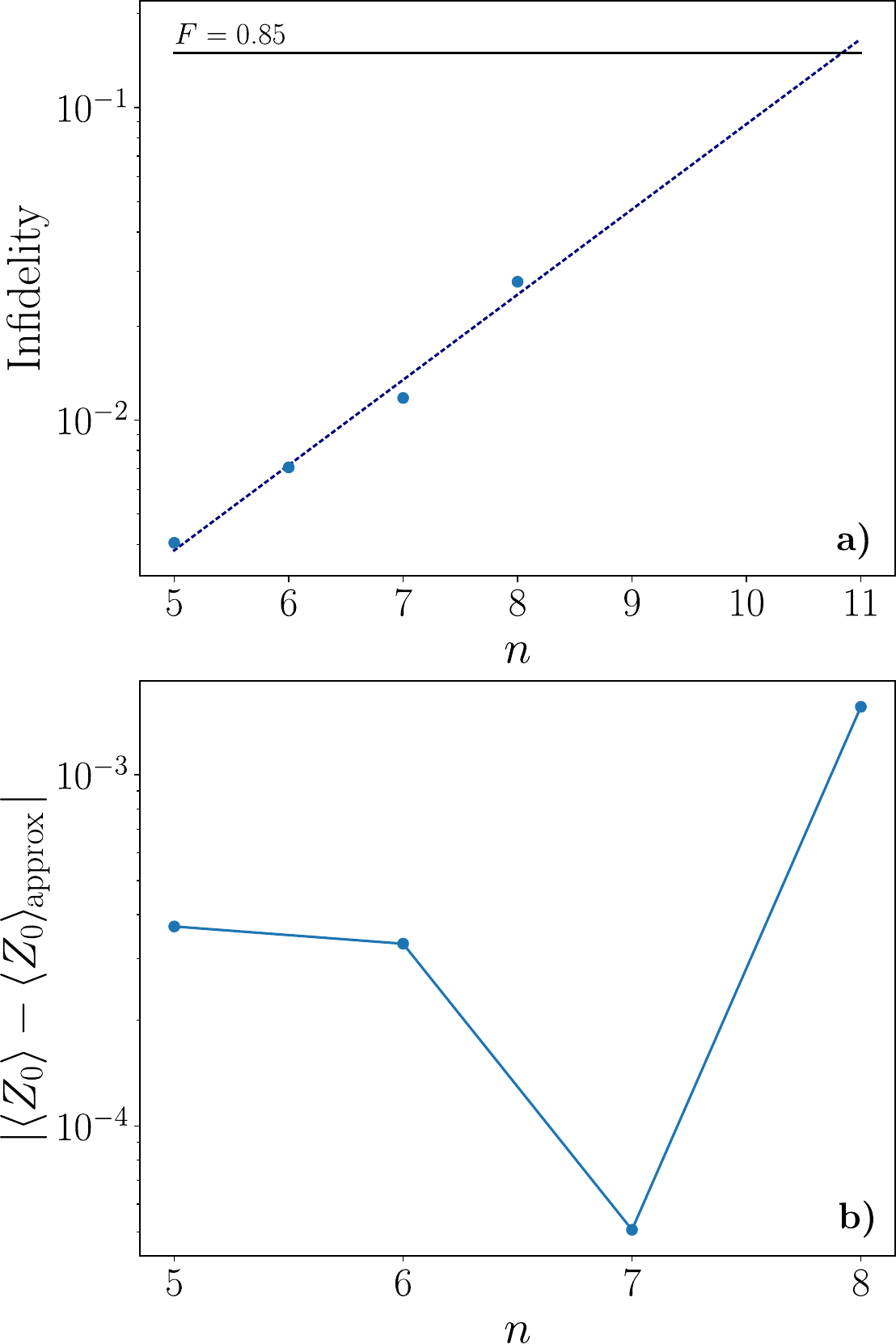} 
\caption{{\bf Scaling of the error mitigation results with the number of qubits $n$.} (a) Infidelity (extrapolated up to $n=11$), and (b) $|{\langle Z_0 \rangle}-{\langle Z_0 \rangle}_\text{approx}|$. The blue dots represent the obtained points. In Figure (a), the blue dotted line represents the infidelity extrapolation, and the black line is the fidelity value $F=0.85$. The simulations were made with $T_1 = 50 \ \mu$s, $p=0.35$ and $g_0 = 1$ MHz.}
\label{Fig: error mitigation 2}
\end{figure}

We can also study the scaling of the infidelity, defined as $1-F$, and the error in the approximation of the expectation value $\langle Z_0 \rangle$ with the number of qubits (Fig.\ \ref{Fig: error mitigation 2}). In Fig.\ \ref{Fig: error mitigation 2}(a), we observe how the fidelity decreases with the number of qubits. Still, using linear regression, we can reach fidelities of approximately 0.85 for up to 11 qubits. When it comes to the computation of $\langle Z_0 \rangle$ (Fig.\ \ref{Fig: error mitigation 2}(b)), the value to approximate varies with each number of qubits as the quantum circuit changes. Thus, we cannot observe a similar scaling for the infidelity, but we do achieve errors of a similar order of magnitude for the different numbers of qubits.

The results obtained in Fig.\ \ref{Fig: error mitigation} show that it is possible to recover the ideal results from applying the bDAQC paradigm by resorting to an adapted zero noise extrapolation. Although we have focused on bDAQC, these results can be extended to sDAQC, as only the zero decoherence limit will be studied in this case. Other error mitigation techniques, such as adapting Clifford Data Regression\ \cite{Czarnik2021}, have been considered. However, these approximations result worse, as we cannot independently correct the error of adding the bDAQC extra interaction. For a more detailed discussion of the error mitigation for bDAQC and concrete results see Appendix~\ref{App}.

\section*{Discussion}

This manuscript aimed to extend the error mitigation techniques to the DAQC paradigm. It first shows an exhaustive platform-based comparison between the DAQC and DQC paradigms to determine if DAQC is a sensible alternative paradigm in the NISQ era. We have illustrated the experiment by implementing the QFT algorithm, considering a noise model that comprises the most relevant noise sources in current superconducting quantum processors. Our simulations conclude that bDAQC is a suitable alternative for NISQ processors. For 3 qubits, the fidelity is about $90\%$ for bDAQC and $85\%$ for sDAQC, while for DQC, it is about $80\%$. As the number of qubits increases, we have observed that the fidelity values obtained for bDAQC still considerably surpass the ones of DQC. However, sDAQC performs worse due to the increasing number of analog blocks affected by the bit-flip. For a higher number of qubits, the noise sources' effect leads to low fidelity values to obtain conclusive results from the output of the experiments.

We have also tested the application of the DAQC paradigm for the QPE algorithm, which has not been used in the DAQC literature to the best of our knowledge. This algorithm is a subroutine in important applications of quantum computing, and hence, it is vital to find the optimum paradigm to minimize the errors. We have tested the performance of this algorithm under a large range of values for the different noise sources, and we conclude that, again, the bDAQC paradigm offers a competitive approximation to the ideal result.

Even though the results demonstrate that bDAQC outperforms DQC under realistic noise sources, the scaling of the impact of noise with the number of qubits and the length of the SQGs is still critical. We have proven that quantum error mitigation techniques can be adapted to the DAQC paradigm with a concrete implementation of one of the most standard techniques, i.e., zero noise extrapolation. This technique enables reducing the effect of noise sources modeled as noise channels, such as decoherence, as well as the inherent banged error. This leads to fidelities around 0.95 for 8 qubits and an error in the approximation of expectation values of order $10^{-3}$, reinforcing the proposal of this paradigm as an appropriate alternative to DQC in the NISQ era.

The results are highly encouraging for the development of NISQ computation architectures. As an outlook, this study can be extended to consider other noise channels, such as correlated noises, to match the performance of different architectures\ \cite{Postler2018,Bermudez2019}, as well as the comparison of our predictions with actual implementations based on transmon or flux qubits. 

\section*{Methods}
\section{Digital-analog quantum computing}\label{sec: DAQC}

In Ref. \cite{Parra2018}, Parra-Rodriguez \textit{et al.} introduced a quantum computation paradigm called DAQC. A DAQC algorithm combines elementary SQGs with high accuracy with current technology, entangling natural interactions among qubits in the processor. In addition to superconducting circuits \cite{Yu2021}, this paradigm of computation can be straightforwardly adapted to other quantum platforms such as trapped ions \cite{Porras2004}, neutral atoms~\cite{Wintersperger2023}, Rydberg atoms \cite{Glaetzle2017}, or nuclear magnetic resonance \cite{Lloyd1993,Cory1998}.

In a superconducting quantum processor, digital blocks are implemented by short microwave pulses, while the analog blocks are derived from the qubit-to-qubit capacitive or inductive coupling. This interaction depends on the type of coupling, the topology of the architecture, or the nature of the qubit, among other aspects. Still, it can be generally expressed as a spin Hamiltonian. In our case, for the sake of simplicity, we have chosen a homogeneous ATA Ising Hamiltonian. Although, in the case that qubits were coupled by an arbitrary inhomogeneous nearest-neighbor Ising Hamiltonian, it is possible to generate an arbitrary ATA Ising Hamiltonian only by employing single-qubit rotations \cite{Galicia2019}
\begin{equation}\label{eq: H_0}
   H_0=H_{\text{int}}=g\sum_{j<k}^N Z^{(j)} Z^{(k)} \quad \rightarrow \quad U_{\text{int}}(t)=e^{i t H_{\text{int}}}, 
\end{equation}
where $g$ is the coupling constant among qubits and $Z^{(i)}$ is the Pauli matrix $\sigma_z^{(i)}$ applied on the $i$-th qubit. However, the DAQC paradigm can be adapted to other emergent Hamiltonians, resulting in a universal paradigm for all.

The homogeneous ATA two-body Ising Hamiltonian is a particular case of the inhomogeneous version,
\begin{equation}
    H_{ZZ}=\sum_{j<k}^N g_{j k}Z^{(j)} Z^{(k)}, \quad\text{with}\quad U_{ZZ}=e^{it_F H_{ZZ}},
\end{equation} 
The coupling constant $g_{jk}$ is different for each pair of qubits. Review the method introduced in Ref. \cite{Parra2018} and its extension in Ref.~\cite{GdA2024} for completeness. Our first goal is to construct an arbitrary inhomogeneous Hamiltonian that combines SQGs and analog blocks. Indeed, we can represent the evolution $U_{ZZ}$ of an $N$-qubit inhomogeneous ATA two-body Ising Hamiltonian by $N(N-1)/2$ time slices $U_\text{int}$ with times $\lbrace t_{jk}\rbrace_{j<k}^N$, where $j$, $k$ are the indices of the application of the $Z$ gates. More specifically, the analog block $U_\text{int}(t_{nm})$ is sandwiched by $X^{(n)} X^{(m)}$ rotations. The map between both pictures is 
\begin{align}\label{eq_ map}
H_{ZZ}&=\sum_{j<k}^N g_{j,k}Z^{(j)} Z^{(k)} \nonumber \\
&=\frac{g}{t_F}\sum_{j<k}^N\sum_{n<m}^N t_{nm} X^{(n)} X^{(m)} Z^{(j)} Z^{(k)} X^{(n)} X^{(m)}.
\end{align}
Let us see how the times $t_{nm}$ must be chosen.
Pauli matrices verify $\lbrace\sigma_i,\sigma_j\rbrace=2\delta_{ij}\mathbbm{I}$ for $i=x,y,z$, so for a term of the type $Z^{(k)} X^{(n)}$, it holds that $Z^{(k)} X^{(n)} =(-1)^{\delta_{kn}}X^{(n)} Z^{(k)}$, and analogously for the other terms. Using this we can write Eq. \eqref{eq_ map} as
\begin{equation}
H_{ZZ}=\frac{g}{t_F}\sum_{j<k}^N\sum_{n<m}^N t_{nm}(-1)^{\delta_{nj}+\delta_{nk}+\delta_{mj}+\delta_{mk}}Z^{(j)} Z^{(k)}.
\end{equation}
This leads to a determined system of $N(N-1)/2$ linear equations, where the unknowns are the time steps of the analog blocks. The value of each time $t_{nm}$ the matrix inversion problem gives 
\begin{equation}
g_\beta=t_\alpha M_{\alpha \beta}\frac{g}{t_F}\quad\rightarrow\quad t_\alpha=(M^{-1})_{\alpha\beta}g_\beta \frac{t_F}{g},
\end{equation}
where $M_{\alpha\beta}$ is a doubly stochastic matrix with elements $\pm 1$, $M_{\alpha\beta}=(-1)^{\delta_{nj}+\delta_{nk}+\delta_{mj}+\delta_{mk}}$. The new indexes $\alpha$ and $\beta$ are introduced to vectorize each pair of indexes $(j,k)$ and $(n,m)$ as
\begin{align}
\alpha=N(n-1)-\frac{n(n+1)}{2}+m, \\
\qquad\beta= N(j-1)-\frac{j(j+1)}{2}+k.
\end{align}
This matrix is invertible $\forall N \in \mathbb{Z}-\lbrace 4 \rbrace$ \cite{LPSS2018}. For a description of the full method for dealing with an arbitrary number of qubits and the negative times problem, we refer the readers to Ref.~\cite{GdA2024}. This work shows a constructive method for obtaining a set of positive times for every problem with classical runtime $O(4^n)$.

\begin{figure}
\centering
\includegraphics[width=0.48\textwidth]{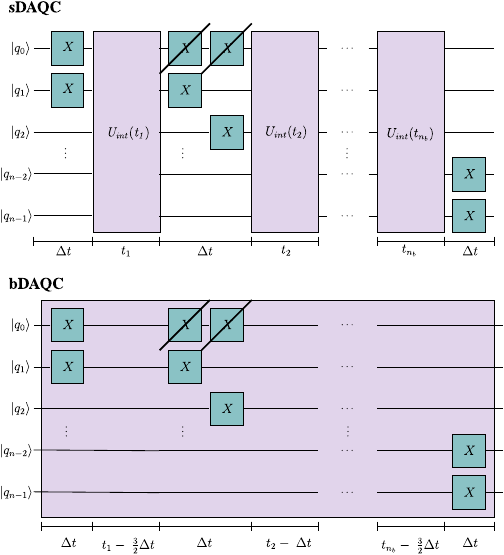} 
\caption{{\bf Scheme of implementing the sDAQC and bDAQC paradigms}. The analog blocks are depicted in purple and represent the interaction given by $U_\text{int}$ in Eq. \eqref{eq: H_0}. The digital blocks are the blue ones, given by $X$ gates. We can observe that in bDAQC, the digital blocks act on top of the analog interaction, while for sDAQC, this interaction is turned on and off, alternating with the digital blocks.} \label{Fig:sDAQC_vs_bDAQC}
\end{figure}

The method described is called stepwise DAQC (sDAQC), which is universal and, therefore, equivalent to DQC. However, it shows the disadvantage that the analog blocks must be switched on and off, which induces critical control errors. Another paradigm is banged DAQC (bDAQC), in which the interaction Hamiltonian is never switched off, and SQRs are applied on top of the analog dynamics. This introduces an inherent error since it does not implement the exact algorithm. Still, when considered experimental errors, they scale up better than sDAQC and DQC \cite{MLSS2020,MCRMCLOSS2021,Headley2020}. This intuitively holds when the application time of SQRs, $\Delta t$, is much smaller than the time scale of the analog blocks. Hence, the error introduced is smaller than the one coming from switching the Hamiltonian in sDAQC on and off. Indeed, experimentally, changing the interaction is not an exact step function, and it takes some time to stabilize. A scheme of implementing both sDAQC and bDAQC is shown in Fig.~\ref{Fig:sDAQC_vs_bDAQC}.

\section{Theoretical description of the effect of noise sources in quantum systems} \label{sec: model}

Quantum computers are open quantum systems, i.e., they cannot be completely isolated from the environment. Thus, they are susceptible to the noise caused by the control and the interaction with the environment. The noise sources generally depend on the experimental setup used to run the quantum experiments. Our theoretical noise treatment will employ the quantum channel formalism, particularly the operator-sum representation. In this formalism, the state of a quantum system is represented by a density matrix $\rho$. The general description of a quantum channel is a linear application, a completely-positive-trace-preserving map $\mathcal{E}$, which maps the initial state $\rho$ into the final one $\rho'$, i.e.,
\begin{equation}
    \rho'=\mathcal{E}(\rho).
\end{equation}
It encloses the system's dynamics change by applying an operator, which describes a physical process \cite{Nielsen2000}. These processes are, for example, the application of a quantum operation on a qubit or the de-excitation of a state due to the interaction with the environment. 

The operator-sum representation is a rigorous mathematical formalism for quantum channels. Let us suppose that we have an open quantum system comprising our system of interest $S$ and the environment $E$, with Hilbert spaces $\mathcal{H}_S$ and $\mathcal{H}_E$, respectively, and $\lbrace\ket{e_k}\rbrace$ is an orthonormal basis for the finite-dimensional Hilbert space corresponding to the environment. Let us assume that the initial state of the environment is $\rho_\text{env}=\dyad{e_0}{e_0}$. For the transformation $U$, the quantum operation can be written as \cite{Nielsen2000}

\begin{align}
    \mathcal{E}(\rho)  &=\text{tr}_{\text{env}}\left[U\left(\rho\otimes\dyad{e_0}{e_0}\right)U^{\dagger}\right] \nonumber \\
      &= \sum_k\bra{e_k}U\left(\rho\otimes\dyad{e_0}{e_0}\right)U^{\dagger}\ket{e_k} 
      = \sum_k E_k\rho E_k^{\dagger}, 
\end{align}
where $E_k\equiv \bra{e_k}U\ket{e_0}$ is an operator acting on the system's subspace. The operators $E_k$ are called Kraus operators\ \cite{Kraus1983} and play a key role in describing the effect of noise sources in quantum systems.

Noise channels are non-unitary quantum channels that reproduce the effect of noise in the state of the system of interest, in our case, the qubits comprising a quantum computer. Let us first focus on single-qubit quantum channels, which only act on one qubit. More concretely, we study the noise channels for superconducting quantum processors' three most common noise sources. These are bit-flip, decoherence, and measurement error.

The bit-flip channel is a model for errors in which the qubit's state is randomly switched with a probability $p$. It is also present in classical computers. The energy that causes the flip of the qubit can be accidentally provided by pulses used to control the qubits or by thermal fluctuations. The Kraus operators of the bit-flip channel are

\begin{align}
    & E_0=\sqrt{1-p}\begin{pmatrix}
1 & 0\\
0 & 1
\end{pmatrix}=\sqrt{1-p}\mathbbm{I},\\
& \quad E_1=\sqrt{p}\begin{pmatrix}
0 & 1\\
1 & 0
\end{pmatrix}=\sqrt{p}X,
\end{align}
where $p$ is the bit-flip probability, $\mathbb{I}$ is the $2\times 2$ identity matrix and $X$ is the $\sigma_x$ Pauli matrix. When the measurement of the system is performed, the qubits might be affected by a bit-flip error. Based on this idea, we will introduce the measurement error as a bit-flip error preceding a perfect measurement operation.

Decoherence is represented through the generalized amplitude-damping channel. This quantum channel accounts for the de-excitation of a two-level system, i.e., the loss of amplitude of the excited state $\ket{1}$ that will decay into the ground state $\ket{0}$ with probability $p$. It can describe different physical phenomena related to energy dissipation, such as the spontaneous emission of a photon and the scattering and attenuation of the state of a photon in a cavity, among others \cite{Nielsen2000}. The Kraus operators of the generalized amplitude-damping channel are

\begin{align}
   & E_0=\sqrt{p}\begin{pmatrix}
1 & 0\\
0 & \sqrt{1-\gamma}
\end{pmatrix}, \quad \\
& E_1=\sqrt{p}\begin{pmatrix}
0 & \sqrt{\gamma}\\
0 & 0
\end{pmatrix}, \\
   & E_2=\sqrt{1-p}\begin{pmatrix}
\sqrt{1-\gamma} & 0\\
0 & 1
\end{pmatrix}, \quad \\
& E_3=\sqrt{1-p}\begin{pmatrix}
0 & 0\\
\sqrt{\gamma} & 0
\end{pmatrix},
\end{align}
where the stationary state of the environment is

\begin{equation}\label{environment}
    \rho_{\text{env}}=\begin{pmatrix}
    p & 0 \\
    0 & 1-p
    \end{pmatrix},
\end{equation}
with $p$ the thermal population of the ground state.

The damping process can be understood as a time-accumulative problem, in which the probability grows with time \cite{Preskill2018_2019}. It can be shown that $\gamma$ can be written as $\gamma=1-e^{-t/T_1}$, where $t$ is the time and $T_1$ is the thermal relaxation time. It describes processes due to the coupling of the qubit to its neighbors, an extensive system in thermal equilibrium at a temperature much higher than the one of the qubit \cite{Nielsen2000}.

Additionally, we also consider the control-related experimental errors described in Refs. \cite{LPSS2018,MLSS2020}. Instead of a quantum-channel approach, we employ quantum trajectories \cite{Carmichael1999}. The error due to SQGs is common to DQC and DAQC, as both use them as a resource to implement quantum algorithms. It simulates a uniform magnetic field noise affecting the qubits. It is given by the parameter $\Delta B$, a random variable from a uniform probability distribution $\Delta B \in\mathcal{U}(1-r_D \Delta t/2,1+r_D \Delta t/2)$, where $\mathcal{U}(a,b)$ stands for a uniform noise distribution with range boundaries $(a,b)$. The deviation ratio is given by $r_D$. This can be modeled as
    \begin{equation}
        e^{i\theta_k Z }\ \rightarrow \ e^{i\theta_k \Delta B Z },
    \end{equation}
where $Z$ stands for any SQG and $\theta_k$ is the rotation angle. 

The noise due to the interaction among qubits is modeled differently for DQC and DAQC. In DQC, any TQG can be decomposed in terms of CZ gates and SQGs. As the CZ gate can be written in terms of SQGs and $ZZ$ interactions of the type described in Eq. \eqref{eq:H_ZZ}, the effects of noise in any TQG can be modeled using the magnetic field noise for SQGs and a Gaussian phase noise for the two-qubit $ZZ$ interaction. This Gaussian phase noise is represented by a random parameter $\varepsilon\in\mathcal{N}(0,\sigma_D)$ affecting the phase of the interaction, where $\mathcal{N}(\mu,\sigma)$ stands for a Gaussian distribution with mean $\mu$ and standard deviation $\sigma$. Its effect on a fixed $\pi/4$ phase interaction, which is usually the interaction of a digital quantum computer, is described by the change
\begin{equation}
        e^{i\frac{\pi}{4}ZZ} \ \rightarrow \ e^{i\frac{\pi}{4}(1+\varepsilon)ZZ}.
\end{equation}
For the DAQC paradigm, a Gaussian coherent noise affecting the time of the analog blocks is introduced to simulate the control error. It produces a switch in the time from $t_{\alpha} \ \rightarrow \ t_{\alpha}+\delta$, where $\delta \in \mathcal{N}(0,r_b\Delta t)$, with $r_b$ the deviation ratio of the time $\Delta t$ required for a single-qubit rotation. This can be implemented as
    \begin{equation}
        e^{it_{\alpha}H_\text{int}} \ \rightarrow \ e^{i(t_{\alpha}+\delta)H_\text{int}},
    \end{equation}
where $H_\text{int}$ is the natural interacting Hamiltonian of the quantum processor.

The parameters $\varepsilon$, $\Delta B$, and $\delta$ must be chosen for each case and individually adapted to the different situations. The Gaussian coherent phase noise has a greater value for sDAQC than for bDAQC, as a result of the effect of switching on and off the interaction in sDAQC, as explained in section \hyperref[sec: DAQC]{Digital-analog quantum computing}. The value of these parameters changes each time a gate is applied to the circuit, as the effects of the noise described can vary in each application.

\section*{Acknowledgements}
The authors thank Juan José García-Ripoll for the valuable discussions. The authors acknowledge financial support from the Basque Government through Grant No. IT1470-22, the Spanish Ramón y Cajal Grant No. RYC-2020-030503-I and the project Grant No. PID2021-125823NA-I00 funded by MCIN/AEI/10.13039/501100011033 and by “ERDF A way of making Europe” and “ERDF Invest in your Future,” as well as from the HORIZON-CL4-2022-QUANTUM-01-SGA project 101113946 OpenSuperQPlus100 of the EU Flagship on Quantum Technologies, and the EU FET-Open projects EPIQUS (899368). Finally, M.S. acknowledges support from the IKUR Strategy under the collaboration agreement between the Ikerbasque Foundation and BCAM on behalf of the Department of Education of the Basque Government. MGdA acknowledges support from the UPV/EHU and TECNALIA 2021 PIF contract call, from the Basque Government through the "Plan complementario de comunicación cúantica" (EXP.2022/01341) (A/20220551), from the Basque Government through the ELKARTEK program, project "KUBIT - Kuantikaren Berrikuntzarako IkasketafTeknologikoa" (KK-2024/00105), and from the Spanish Ministry of Science and Innovation under the Recovery, Transformation and Resilience Plan (CUCO, MIG-20211005). P.G.M. acknowledges support from  MCIU/AEI/FEDER, CSIC Research Platform PTI-001, the Proyecto Sin\'ergico CAM 2020 Y2020/TCS-6545 (NanoQuCoCM), and MCIN/AEI/10.13039/501100011033 and "FSE invierte en tu futuro" through an FPU Grant. FPU19/03590. 

\section*{Data availability}

The datasets generated and analyzed during the current study are available from the corresponding author upon reasonable request.

\section*{Code availability}\label{sec: code}

The code created for this study is available from the corresponding author upon reasonable request. Code for the simulation of DAQC circuits with noise and transpilation of digital circuits to DAQC is available in \url{https://github.com/NQUIRE-Center/DAQC_simulator}. 

\section*{Author contributions}

P.G.M. developed the original code for the QFT and its noisy version, created Figs 1, 5, 6, and 7, and wrote the main manuscript. A.M. supervised the development of the code, helped with the theoretical background, and did Figs. 2,3. M.G.A. did the code for the QPE and for Figs. 3, 4. M. S. provided the seminal ideas and theoretical background and supervised the article through all stages. All authors have carefully reviewed the manuscript.

\section*{Competing interests}

The authors declare no competing interests.

\appendix

\section{\uppercase{Error mitigation}}\label{App}

In the subsection \hyperref[sec: error mitigation]{Error mitigation in DAQC}, we demonstrated how to apply error mitigation to the DAQC paradigm. This appendix includes a more detailed explanation of the procedure, the values chosen for these simulations, and the numerical results obtained.

First, to have a more significant effect on the noise sources in our system, we choose $g_0 = 1$ MHz and only consider the impact of decoherence using the thermal relaxation channel with $T_1 = 50 \ \mu$s and $p=0.35$. We opt for a $8$-qubit initial state{$\ket{\psi_0}=\sin\pi/4\ket{\text{W}_8}+\cos\pi/4\ket{\text{GHZ}_8}$ and apply the bDAQC QFT circuit on it to study its fidelity and expectation value $\langle Z_0 \rangle$ for different values of $\Delta t_{\text{SQG}_{i,j}} = b_i/g_j$, with $b_i=[1/50,1/100,1/150,1/200,1/250]$ and $g_j = a_j g_0$, $a_j=[0.94,0.97,1.00,1.03,1.07]$.

The procedure is as follows. For each value of $b_i$, we perform a zero noise linear extrapolation with the time associated to each $g_j$ as the noise parameter, leading to the zero decoherence limit for each $b_i$, i.e., $\Delta t_{\text{SQG}_i} = b_i/g_0$, for the fidelity and the expectation value $\langle Z_0 \rangle$ (see Tables \ref{tab: step1 fidelity}-\ref{tab: step1 expectation value}). Using the zero decoherence limit values of these figures of merit, we can move on to step two of the error mitigation for bDAQC, which is used to remove the intrinsic error associated with this paradigm. We use different extrapolation techniques and obtain the results in Tables \ref{tab: step2 fidelity}-\ref{tab: step2 expectation value}.
\begin{table}[!t]
\centering
\begin{tabular}{|c|c|c|c|c|c|} \hline
  & $b_0$ & $b_1$ & $b_2$ & $b_3$ & $b_4$ \\ \hline
\multicolumn{1}{|c|}{$g_0$} & 0.2807 & 0.4982 & 0.5648 & 0.5918 & 0.6052 \\ \hline
\multicolumn{1}{|c|}{$g_1$} & 0.2848 & 0.5054 & 0.5729 & 0.6003 & 0.6138 \\ \hline
\multicolumn{1}{|c|}{$g_2$} & 0.2887 & 0.5122 & 0.5806 & 0.6083 & 0.6221 \\ \hline
\multicolumn{1}{|c|}{$g_3$} & 0.2924 & 0.5187 & 0.5880 & 0.6160 & 0.6300 \\ \hline
\multicolumn{1}{|c|}{$g_4$} & 0.2972 & 0.5270 & 0.5973 & 0.6257 & 0.6398 \\ \hline
\multicolumn{1}{|c|}{\begin{tabular}[c]{@{}l@{}}Zero decoherence\\ \centering limit\end{tabular}} & 0.4187 & 0.7398 & 0.8370 & 0.8761 & 0.8953 \\ \hline
\multicolumn{1}{|c|}{Ideal} & 0.4517 & 0.7957 & 0.8994 & 0.9409 & 0.9613 \\ \hline
\end{tabular}
\caption{\textbf{Fidelity results for the first step of the zero noise extrapolation for bDAQC (zero decoherence).}}
\label{tab: step1 fidelity}
\end{table}
\newpage

\begin{table}[!t]
\centering
\begin{tabular}{|c|c|c|c|c|c|}\hline
 & $b_0$ & $b_1$ & $b_2$ & $b_3$ & $b_4$ \\ \hline
\multicolumn{1}{|c|}{$g_0$} & 0.0665 & 0.1518 & 0.1753 & 0.1845 & 0.1891 \\ \hline
\multicolumn{1}{|c|}{$g_1$} & 0.0669 & 0.1529 & 0.1766 & 0.1859 & 0.1905 \\ \hline
\multicolumn{1}{|c|}{$g_2$} & 0.0672 & 0.1539 & 0.1778 & 0.1872 & 0.1918 \\ \hline
\multicolumn{1}{|c|}{$g_3$} & 0.0675 & 0.1549 & 0.1789 & 0.1884 & 0.1931 \\ \hline
\multicolumn{1}{|c|}{$g_4$} & 0.0679 & 0.1561 & 0.1804 & 0.1899 & 0.1946 \\ \hline
\multicolumn{1}{|c|}{\begin{tabular}[c]{@{}l@{}}Zero decoherence\\ limit\end{tabular}} & 0.0777 & 0.1874 & 0.2174 & 0.2291 & 0.2348 \\ \hline
\multicolumn{1}{|c|}{Ideal} & 0.0750 & 0.1882 & 0.2190 & 0.2310 & 0.2369 \\ \hline
\end{tabular}
\caption{\textbf{$\boldsymbol{\langle Z_0 \rangle}$ results for the first step of the zero noise extrapolation for bDAQC (zero decoherence).}}
\label{tab: step1 expectation value}
\end{table}

\begin{table}[!t]
\centering
\begin{tabular}{|c|c|c|c|c|}\hline
                                                                 Ideal  & Linear  & Quadratic & Cubic  & Richardson \\\hline
1 & 0.9723 & 0.9398 & 0.9304 & 0.9723  \\\hline

\end{tabular}
\caption{\textbf{Fidelity results for the second step of the zero noise extrapolation for bDAQC (zero $\boldsymbol{\Delta t_\text{SQG}}$).}}
\label{tab: step2 fidelity}
\end{table}

\begin{table}[!t]
\centering
\begin{tabular}{|c|c|c|c|c|}\hline
                                                                 Ideal  & Linear  & Quadratic & Cubic  & Richardson \\\hline
0.2500 & 0.2578 & 0.2484 & 0.2474 & 0.2468  \\\hline

\end{tabular}
\caption{\textbf{$\boldsymbol{\langle Z_0 \rangle}$ results for the second step of the zero noise extrapolation for bDAQC (zero $\boldsymbol{\Delta t_\text{SQG}}$).}}
\label{tab: step2 expectation value}
\end{table}

\end{document}